\documentstyle[aps,prl,epsf]{revtex}
\begin{document}
% for two column  activate the line below...                
\twocolumn[\hsize\textwidth\columnwidth\hsize\csname@twocolumnfalse\endcsname
\title{Crossing lattices, vortex chains, and angular dependence of melting line in 
layered superconductors} 
\author{A.E.Koshelev} 
\address{Materials Science Division, Argonne National Laboratory, 
Argonne, Illinois 60439}
\date{\today} 
\maketitle
\tighten
\begin{abstract}
We investigate vortex structure in very anisotropic layered 
superconductors at fields tilted with respect to the {\em c}-axis.  We 
show that even a small in-plane field does not tilt the vortex lattice 
but, instead, penetrates inside the superconductor in the form of 
Josephson vortices (JVs).  At high c-axis magnetic field the phase 
field of the JV is built up from the phase perturbations created by 
displacements of pancake vortices.  The crossing-lattices ground state 
leads to linear dependencies of the melting field and melting 
temperature on the in-plane field, in agreement with recent 
experimental observations.  At small fields stacks of JVs accumulate 
additional pancake stacks creating vortex rows with enhanced density.  
This mechanism explains the mixed chains-lattice state observed by 
Bitter decorations.
\end{abstract}
% for two column  activate the line below... 
%\vspace*{-10pt}
\twocolumn 
\vskip.2pc] 
\narrowtext 

Significant progress has been achieved recently in understanding of 
vortex states and melting transition in layered superconductors with 
very weak Josephson coupling such as Bi$_{2}$Sr$_{2}$CaCu$_{2}$O$_{x}$ 
(BSCCO).  A magnetic field applied perpendicular to the layers 
penetrates such superconductors in the form of pancake vortices (PVs) 
\cite{pancakes}.  PVs in different layers are coupled weakly via the 
Josephson and magnetic interactions and form aligned stacks at low 
fields and temperatures (PV stacks).  The magnetic nature of coupling 
leads to an unusual melting scenario.  In contrast to superconductors 
with strong Josephson coupling, both crystalline order and alignment 
of PVs are destroyed at the melting point leading to almost 
independent two-dimensional vortex liquids in the layers 
\cite{Blatter-Magn}.

\begin{figure}
\epsfxsize=3.2in \epsffile{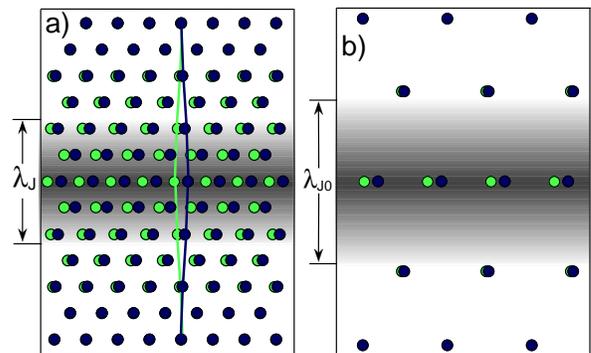} \caption{JV coexistent with 
the lattice of PVs.  Shaded area shows the JV core.  Pancake positions 
in the two central layers are shown.  a) At large fields the JV core 
contains a large number of pancake rows.  The phase field of the JV is 
determined by displacements of the PVs.  b) At small fields the JV 
core captures only one pancake row.  Due to the interaction energy 
between the PV stacks and JVs, pancake density in this row is higher 
than in neighboring rows (mixed chains-lattice state).}
\label{Fig-JVpanc}
\end{figure}

A rich variety of vortex structures were proposed for the magnetic 
field tilted with respect to the {\em c}-axis, such as the kinked lattice 
\cite{BLK,Feinberg93}, crossing lattices of Abrikosov and Josephson 
vortices (JVs) (combined lattice) \cite{BLK}, tilted vortex chains 
\cite{chains}, etc.  Obviously, the thermodynamics of the melting 
transition at tilted fields is determined by the nature of the ground 
state.  In moderately anisotropic superconductors an in-plane field 
simply tilts the vortex lattice inside the superconductor.  This 
scenario does not work in the case of very weak Josephson coupling 
when the interlayer coupling is predominantly magnetic.  The in-plane 
magnetic field interacts with the PVs only via the 
Josephson coupling.  On the other hand, the PVs are mainly aligned by 
magnetic coupling, therefore a homogeneous tilt of the lattice costs 
magnetic energy.  To save the magnetic coupling energy it is more 
favorable for the in-plane field to penetrate inside the 
superconductor in the form of JVs as in the Meissner state.  The 
ground state of the system is then given by the dilute lattice of the 
JVs coexisting with the dense lattice of the PVs (crossing lattices or 
combined lattice).  This ground state was first proposed for the 
opposite case of a small {\em c}-axis field \cite{BLK} when the dilute 
pancake lattice coexists with the dense Josephson lattice.  The JV in 
weakly coupled superconductors has a very wide nonlinear core (the 
region where the interlayer phase difference is large).
%In contrast to Abrikosov vortex JV does not have 
%normal core.  The nonlinear core of JV is defined as a region within which the 
%interlayer phase difference is large.  This region is very broad when the 
%coupling is weak.  
This means that even at moderate {\em c}-axis fields the JV core contains 
many PVs (see Fig.~\ref{Fig-JVpanc}a).  We show that in this situation the 
phase field of the JV is built up from the phase perturbations created by 
pancake displacements.  Such a JV has a smaller core size and smaller 
energy as compared to the ordinary JV. In the crossing-lattices state the 
free energy depends almost linearly on the in-plane field.  As a 
consequence, the melting temperature and melting field of the pancake 
lattice also depend on the in-plane field in a linear way in agreement with 
recent experiments \cite{Schmidt,Ooi}.  At high in-plane field the JV cores 
start to overlap.  In the regime of strongly overlapping JVs the in-plane 
field produces only weak zigzag deformations in the PV lattice 
\cite{Bul-zigzag}.  These deformations have a very little influence on the 
phase distribution and can be treated perturbatively, in contrast to 
isolated JV.

At very small {\em c}-axis fields the JV core contains only one row of PVs.  
In this case displacements of the PVs have a very little influence on the 
JV structure.  These displacements, however, produce a finite interaction 
energy between the PV stack and the JV. Due to a very high anisotropy the 
JV lattice consists from densely packed stacks along {\em c}-axis (JV 
stacks) separated by relatively large distances.  PV stacks crossing JV 
stacks have smaller energy as compared to other stacks and it would be 
favorable to add extra PV stacks to the rows located on the JV stacks.  
However this makes the PV lattice defective and increases its energy.  On 
the other hand, when the distance between PVs $a$ is much larger than the 
London penetration depth $\lambda$, the energy of defect in the lattice is 
exponentially small.  This means that at a certain field there is a phase 
transition from the weakly deformed triangular lattice to the defective 
lattice.  In the latter state PV rows located on the JV stacks will have a 
larger density of PVs than other rows (see Fig.~\ref{Fig-JVpanc}b).  This 
picture provides a natural explanation for the mixed chain-lattice state 
observed by Bitter decorations in BSCCO \cite{Bolle91,Grig95}.  Initially 
these chains were attributed to the chains of tilted vortices, which were 
predicted on the basis of the anisotropic London theory 
\cite{chains}.  Further detailed investigations \cite{Grig95} indicated 
that this interpretation does not agree with experiment.  On the other hand 
the interpretation based on the crossing lattices
%first proposed by Huse 
\cite{Huse} qualitatively agrees with the experiment.  We 
will calculate the crossing energy of the JV and PV stack which allow 
to obtain a quantitative criterion for the transition to the ``chain 
state''.

%\section{Josephson vortex in presence of pancake lattice}
%
First, we consider a structure of an isolated JV. In the Meissner state 
the JV is represented by a region of a singular phase 
distribution \cite{ClemJV,Kinkwalls}. The phase difference between two 
central layers changes from $0$ to $2\pi $ within the nonlinear core 
region.  The core size $\lambda_{J0}$ and the JV energy 
$\varepsilon_{J0}$ are determined by the balance condition between 
the in-plane phase stiffness energy, $(J/2)(\nabla \phi _{n})^{2}$, 
and the Josephson energy, $-E_{J}\cos (\phi _{n,n+1})$, 
\cite{Kinkwalls}
\[
\lambda _{J0}\approx \sqrt{J/E_{J}}=\gamma s,\;\;\;
\varepsilon _{J0}=\pi \sqrt{JE_{J}}\left( \ln (\lambda /s)+1.55\right)
\]
where $\phi_{n}({\bf r})$ is the phase of the order parameter in the 
$n$-th layer, $\phi_{n,n+1}=\phi_{n+1}-\phi_{n}-\left(2\pi s/\Phi 
_{0}\right) A_{z}$ is the gauge invariant phase difference, $\lambda$ 
is the London penetration depth, $\gamma$ is the anisotropy parameter, 
$s$ is the interlayer spacing, $J=s\Phi_{0}^{2}/\pi(4\pi 
\lambda^{2})$, and $E_{J}=J/(\gamma s)^{2}$.

We consider now a JV in the presence of a pancake lattice with lattice 
parameter $a\ll \lambda_{J0}$ (see Fig.~\ref{Fig-JVpanc}a) for the 
case of very weak coupling, $\lambda_{J0}> \lambda$.  If we put the JV 
into the PV lattice, its in-plane supercurrents will displace the PVs 
from their equilibrium positions.  Lattice displacements induce extra 
phase variations, which add to the usual regular phase and renormalize 
the JV structure.  We will show that at large $B_{z}$ the vortex phase 
has a smaller stiffness as compared to the regular phase and gives a 
dominating contribution to the phase perturbations.
%Lattice displacements ${\bf u}_{n}({\bf r})$ are rigidly coupled to the phase 
%perturbations.  
%As a consequence, lattice deformations strongly structure of the JV. 
%In general, phase field of the JV consist of vortex and regular contributions.  
The coarse-grained vortex phase $\phi_{nv}(y)$, created by the 
unidirectional shear lattice displacement in the $n$-th layer, $u_{nx}(y)$, 
is determined by $\nabla_{y}\phi_{nv}=2\pi n_{v}u_{nx}$ with 
$n_{v}=B_{z}/\Phi_{0}$.  The lattice deformations produce the phase sweeps 
between the opposite sides of the deformed region 
$\delta_{y}\phi_{nv}\equiv\phi_{nv}(\infty)-\phi_{nv}(-\infty)$ related to 
$u_{nx}(y)$ as $\delta_{y}\phi_{nv}=2\pi 
n_{v}\int_{-\infty}^{\infty}u_{nx}(y)dy$. If we neglect the regular phase, 
these phase sweeps are given by $\delta_{y}\phi_{nv}=\pi{\rm sign}(n)$ in 
the isolated JV. The vortex phase deformations in the JV are large and can 
not be treated perturbatively.
%This is in contrast to the case of dense Josephson lattice considered in 
%Ref.~\onlinecite{Bul-zigzag}.

To consider a structure of the JV quantitatively we calculate the 
phase stiffness due to the lattice deformations.  This phase stiffness is 
mainly determined by the shear part of the elastic energy, which for 
dominating magnetic coupling can be written as \cite{KoshKes93,Horovitz}
\begin{equation}
{\cal E}_{v}= \frac{1}{2}\int \frac{d{\bf k}}{(2\pi)^{3}} \left(
C_{66}k_{\perp }^{2}+U_{44}(k_{z})\right) \left| {\bf u}\right| ^{2}
\label{ShearEn}
\end{equation}
where $C_{66}=\frac{B_{z}\Phi _{0}}{(8\pi \lambda )^{2}}$ and 
$$U_{44}(k_{z})\equiv C_{44}(k_{z})k_{z}^{2}\approx \frac{B_{z}\Phi 
_{0}}{2(4\pi )^{2}\lambda ^{4}}\ln 
\left(1+\frac{r_{cut}^{2}}{k_{z}^{-2}+r_{w}^{2}}\right),$$ with 
$r_{cut}=\min (a,\lambda )$ and $r_{w}\approx u_{1x}(0)>s$.  Using the 
relation between the phase perturbation and the lattice displacements, 
$\phi _{v}\left( {\bf k}\right) =2\pi in_{v}u/k_{\perp }$, we rewrite 
the elastic energy in terms of phase as
\begin{equation}
{\cal E}_{v}=\int \frac{d{\bf k}}{(2\pi)^{3}} \frac{J(B_{z},{\bf 
k})}{2s}k_{\perp }^{2}\left| \phi _{v}({\bf k})\right| ^{2},
\label{ShearPhStiff}
\end{equation}
with the effective phase stiffness $J(B_{z},{\bf k})$, 
$J(B_{z},{\bf k})=$ $\frac{s\left(C_{66}k_{\perp 
}^{2}+U_{44}\right)}{\left( 2\pi n_{v}\right) ^{2}}$.  In the range 
$k_{\perp}< 2/\lambda$ and $k_{z}>1/r_{w}$ the phase stiffness is 
${\bf k}$-independent
\[
J(B_{z})\approx \frac{sU_{44}}{(2\pi n_{v})^{2}} 
=J\frac{B_{\lambda}}{B_{z}}, \; \; B_{\lambda}\equiv\frac{\Phi 
_{0}}{4\pi \lambda ^{2}}\ln \frac{r_{cut}}{r_{w}}
\]
i.e., magnetic field reduces the phase stiffness approximately by the 
factor $B_{\lambda}/B_{z}$.  This justifies neglect of the regular 
phase in the region $B_{z}\gg B_{\lambda}$.  The phase stiffness 
energy (\ref{ShearPhStiff}) has to be supplemented by the Josephson 
energy.  Thermal motion of the PVs induces the fluctuating phase 
$\tilde{\phi}_{n,n+1}$ and suppresses the effective Josephson energy, 
$E_{J}\rightarrow {\cal C}E_{J}$ where ${\cal C}\equiv \left\langle 
\cos \tilde{\phi}_{n,n+1}\right\rangle $.  Therefore the total energy 
in terms of $\phi _{v}$ is given by
%\begin{equation}
\[
{\cal E}=\sum_{n}\int d{\bf r}\left[ 
\frac{J(B_{z})}{2}(\nabla \phi_{vn}) ^{2}
-{\cal C}E_{J}\cos \left(\phi_{vn,n+1}\right) \right].  
\]
%\label{PhaseEn}
%\end{equation}
It has essentially the same form as in the Meissner state except for 
values of the phase stiffness and Josephson energy.  Using known 
structure of an ordinary JV \cite{ClemJV,Kinkwalls} we 
obtain for the core size and energy of the JV:
\begin{equation}
\lambda _{J}(B_{z})= \lambda _{J0}\sqrt{\frac{B_{\lambda}}{{\cal 
C}B_{z}}},\;\;\;
\varepsilon_{J}(B_{z}) \approx \varepsilon_{J0} 
\sqrt{\frac{{\cal C}B_{\lambda}}{B_{z}}}. 
\label{enJ}
\end{equation}
The JV energy is reduced by the $c$-axis field, due to renormalizations of 
the phase stiffness and Josephson energy.  
%Note that the logarithmic divergency in $\varepsilon_{J}(B_{z})$ is
%cut at large distances at $z\approx r_{cut}$ due to crossover in $k_{z}$	
%dependence of the tilt stiffness $U_{44}$ which makes the phase				
%stiffness vanish at small $k_{z}$ ($U_{44}\propto \ln						
%(k_{z}r_{cut})$ at $k_{z}\gtrsim 1/r_{cut}$ and $U_{44}\propto				
%k_{z}^{2}$ at $k_{z} < 1/r_{cut}$).					
The maximum lattice displacement $r_{w}$ and lattice deformation 
$u_{xy}$ in the vortex core can be estimated as $r_{w} \approx 
a\lambda/\lambda_{J0}$ and $u_{xy}\approx a\lambda/\lambda _{J0}^{2}$.
%Condition for using the elasticity theory $u_{xy}\ll 1$ is satisfied already at 
%very small fields $a < \lambda _{J0}^2/\lambda$.                  

%\section{Josephson lattice}
%
Interactions between the JVs are given by the anisotropic London 
theory with the anisotropy ratio $\gamma_{\rm eff}=\gamma 
\sqrt{B_{\lambda}/(B_{z}{\cal C})}$.  At finite density the JVs form a 
triangular lattice, which is stretched along the layers with the ratio 
$\gamma_{\rm eff}$, i.e., the JVs form stacks along the {\em c}-axis 
with the period $c_{z}=\sqrt{\beta \Phi_{0}/(\gamma_{\rm eff}B_{x})}$ 
and these stacks are separated by the distance 
$c_{y}=\sqrt{\gamma_{\rm eff}\Phi_{0}/(\beta B_{x})}$.  Within the 
anisotropic London theory two states with $\beta= 2\sqrt{3}$ and 
$\beta=2/\sqrt{3}$ have the same energy \cite{CDK88}.  These states 
correspond to stretching of the regular triangular lattice with the 
closed-packed direction oriented along the layer and along the {\em c}-axis.  
The lattice energy $F_{JL}$ is given by
%calculated in the same way as the 
%energy of the vortex lattice in anisotropic superconductors,
\begin{equation}
F_{JL}(B_{z},B_{x})=\frac{B_{x}^{2}}{8\pi}+\frac{B_{x}}{\Phi 
_{0}}\frac{\varepsilon _{0}}{ \gamma }
\sqrt{\frac{B_{\lambda}{\cal C}}{B_{z}}}\ln 
\frac{c_z}{s}.
\label{FJ}
\end{equation}
%The difference from usual situation appears again due to 
%renormalization of the phase stiffness and Josephson energy.  
We now compare this energy with the energy change corresponding to the 
homogeneous tilt of the lattice $\delta F_{tilt}$, which in 
magnetically coupled superconductors is determined by the tilt modulus 
$C^{(0)}_{44}$ at $k_{z}=0$\cite{KoshKes93,Horovitz}
\begin{equation}
\delta F_{tilt}=\frac{C^{(0)}_{44}}{2}\left(\frac{B_{x}}{B_{z}}\right)^{2},\;
C^{(0)}_{44}=\frac{B_{z}^{2}}{8\pi}+3.68\frac{\Phi_{0}^{2}}{(4\pi 
\lambda)^{4}}.
\label{dFtilt}
\end{equation} 
Comparing Eqs.~(\ref{FJ}) and (\ref{dFtilt}) we see that the crossing 
lattices is a favorable state when the tilt angle of the field 
$\theta \approx B_{x}/B_{z}$ exceeds the critical value
\begin{equation}
\theta_{0}=\frac{6.8}{\gamma}\sqrt{\frac{4\pi\lambda^{2}B_{z}{\cal 
C}}{\Phi_{0}}\ln \frac{r_{cut}}{r_{w}}}\ln\frac{c_{z}}{s}.
\label{theta0}
\end{equation}
At $\theta=\theta_{0}$ the system experiences a first order phase 
transition from the homogeneously tilted lattice to the crossing 
lattices of  PVs and  JVs.  Taking $\gamma =500$ and $\lambda 
=200$nm$/\sqrt{1-(T/T_{c})^{2}}$ at $B_{z}=130$ G and $T=70$ K we 
obtain $\theta_{0}\approx 3^{\circ}$.

%\section{Influence of Josephson vortices on melting of the pancake lattice}
%
To derive the shift of the melting point by the JV lattice we write
the thermodynamic condition for melting 
\[
F_{cr}(B_{zm})+\delta F_{cr}(B_{zm},B_{x})=F_{l}(B_{zm})+\delta
F_{l}(B_{zm},B_{x}),
\]
where $F_{cr}(B_{z})$ and $F_{l}(B_{z})$ are the free energies of 
the crystal and liquid states for field along the $c-$axis, and $\delta 
F_{cr}(B_{z},B_{x})$ and $ \delta F_{l}(B_{z},B_{x})$ are corrections 
due to $B_{x}$.  For small $B_{x}$ we can write 
$B_{zm}(B_{x})=B_{m0}+\delta B_{m}(B_{x})$ and expand 
$F_{cr}(B_{zm})-F_{l}(B_{zm})$ with respect to $\delta B_{m}(B_{x})$, 
$F_{cr}(B_{zm})-F_{l}(B_{zm})\approx \delta B_{m}(B_{x}) \Delta M$, 
where $\Delta M=M_{l}-M_{cr}>0$ is the magnetization jump at the 
melting point.  This gives 
\begin{equation}
\delta B_{m}(B_{x})\approx -\Delta F(B_{m0},B_{x})/\Delta M
\label{deltaBm}
\end{equation}
where $\Delta F(B_{m0},B_{x})=\delta F_{cr}(B_{m0},B_{x})-\delta 
F_{l}(B_{m0},B_{x})$.  
% In the same way we can relate the shift of the 
% melting temperature $\delta T_{m}(B_{x})$ with the entropy jump 
% $\Delta S$ at the melting point per single PV,
% \begin{equation}
% \delta T_{m}(B_{x})\approx -s\Phi _{0}\Delta F(B_{m0},B_{x})/
% (B_{m0}\Delta S).  
% \label{deltaTm}
% \end{equation}
Similarly, we can relate the shift of the melting temperature $\delta 
T_{m}(B_{x})$ with the melting entropy jump $\Delta S$ per single PV, 
$\delta T_{m}(B_{x})\approx -s\Phi _{0}\Delta F(B_{m0},B_{x})/ 
(B_{m0}\Delta S).  $

\begin{figure}
\epsfxsize=3.2in \epsffile{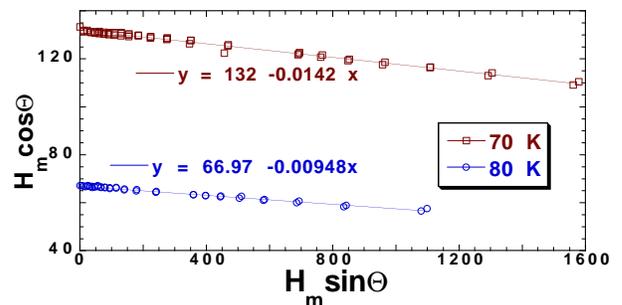} 
\caption{Dependence of the $c$-component of the melting field on the 
in-plane field for BSCCO at $T=70$K and $80$K.  The data are taken from 
Ref.~\protect\onlinecite{Schmidt}.  For both cases a linear dependence 
is clearly observed indicating the crossing-lattices ground state.}
\label{Fig-Schmidt}
\end{figure}

$\delta F_{cr}(B_{m0},B_{x})$ is determined by the energy of the Josephson 
lattice (\ref{FJ}) (we estimated that entropic correction to $F_{JL}$ is 
relatively weak).  In magnetically coupled superconductors the melting 
transition is accompanied by strong misalignment of PVs in the different 
layers \cite{Blatter-Magn}.  This means that the Josephson coupling in the 
liquid state is strongly suppressed and the influence of the in-plane 
magnetic field in the liquid is much weaker than in the crystal, $\Delta 
F(B_{m0},B_{x})\approx F_{JL}(B_{m0},B_{x})- B_x^2/(8\pi)\propto B_x$.  
From Eqs.\ (\ref{FJ}) and (\ref{deltaBm}) we obtain that the shift in 
$B_{mz}$ is approximately linear with $B_{x}$ and
\begin{equation}
\frac{\partial B_{m}}{\partial B_{x}}=-\frac{\Phi _{0}}{4\pi \lambda 
^{2}\gamma \Delta B}\sqrt{\frac{B_{\lambda}{\cal C}}{B_{z}} }\ln 
\frac{c_z}{s},\; \;  \Delta B\equiv 4\pi \Delta M.  
\label{dBmdBx}
\end{equation}
%In a similar way we obtain the derivative of $T_{m}$ with respect to $B_{x}$ 
%\begin{equation}
%\frac{\partial T_{m}}{\partial B_{x}}=-\frac{s\varepsilon _{0}}{\gamma
%B_{z}\Delta S}\sqrt{\frac{\Phi _{0}{\cal C}}{4\pi \lambda ^{2}B_{z}}\ln 
%\frac{r_{cut}}{r_{w}}}\ln \frac{b_z}{s}  \label{dTmdBx}
%\end{equation}
The linear dependence holds until the JV cores start to overlap at 
$B_{x}\approx \gamma \sqrt{B_{\lambda}B_{z}/{\cal C}}$.  In clean 
superconductors the linear dependence also breaks down at large enough 
$B_{z}$ when the decoupling transition occurs within the 
crystalline phase \cite{decoupl}.  The melting field of the tilted vortex lattice in 
moderately anisotropic superconductors depends quadratically on the 
in-plane field \cite{BlatterAnis} in agreement with experiment 
\cite{AnisExp}.  Therefore the linear dependence is an indication of the 
crossing-lattices state.  The angular dependence of the melting field was 
measured in Ref.~\onlinecite{Schmidt}.  Fig.~\ref{Fig-Schmidt} shows data 
from Fig.~4 of this paper for T$=70$K and $80$K replotted in the 
coordinates $H_{mz}=H_{m}\cos \theta$ vs $H_{m\parallel}=H_{m}\sin \theta$.  
In both cases the linear dependence $H_{mz}(H_{m\parallel})$ is clearly 
observed.  The same linear behavior of the melting field has been reported 
recently by Ooi {\em et al.}\cite{Ooi}.  For $T=70$K we have $B_{m}=133$G, 
$\Delta B=0.35$G, and $\partial B_{m}/\partial B_{x}=-0.014$.
%On the other hand, taking $\lambda =2000\AA /\sqrt{1-(T/Tc)^{2}}$ and 
%$\gamma =500$ and neglecting logarithmic factors 
Using the same parameters as before we obtain from Eq.~(\ref
{dBmdBx}) an estimate $\partial B_{m}/\partial B_{x}\approx -0.03$, 
in reasonable agreement with experiment.

%\section{Small fields. Vortex chains}
%
We now consider the case of small {\em c}-axis fields, $B_{z}\ll 
\Phi_{0}/(\gamma s)^{2}$.  In this case the JV core contains only one row 
of PVs (see Fig.~\ref{Fig-JVpanc}b).  Displacements of PVs do not strongly 
influence the JV structure but create a small pinning energy for the PV 
stack by the JV. Due to this energy one can expect at some field a phase 
transition from the weakly deformed triangular lattice to the ``chain 
state'' represented by regions of good lattice intercepted by the vortex 
chains located on the JV stacks.  In the ``chain state'' the degeneracy in the 
JV lattice \cite{CDK88} is broken and the state with smaller $c_{z}$ 
corresponding to $\beta=2/\sqrt{3}$ has smaller energy.  The distance 
between the chains is simply given by the distance between the JV stacks 
$c_{y}=\sqrt{\sqrt{3}\gamma \Phi_{0}/(2B_{x})}$ \cite{Huse}.  For 
$B_{x}\approx 30$G and $\gamma=200$ this gives $c_{y}\approx 10$$\mu$m, 
which roughly agrees with the distance between the chains observed in the 
decoration experiments \cite{Bolle91,Grig95}.  We now obtain the criterion 
for the transition to the ``chain state''.

First, consider one PV stack crossing one JV located between the 
layers 0 and 1.  The crossing energy $E_{\times}$ appears due to 
displacements of PVs $u_{n}$ under the action of the in-plane currents 
$j_{n}$ induced by the JV
\begin{equation} 
E_{\times}\approx\sum_{n}\left(\frac{U_{44}}{2n_{v}}u_{n}^2
-\frac{s\Phi _{0}}{c}j_{n}u_{n}\right)
\label{EJv} 
\end{equation} 
with $j_{n}\approx \frac{c\Phi _{0}}{8\pi ^{2}\lambda 
^{2}}\frac{C_{n}}{(n-1/2)\gamma s}$, where $C_{n}\rightarrow 1$ at 
$n\rightarrow \infty $.  For estimates we neglected the weak 
logarithmic $k_{z}$-dependence of $U_{44}(k_{z})$.  Minimization of 
(\ref{EJv}) with respect to $u_{n}$ gives
\[ 
u_{n}\approx \frac{2C_{n}\lambda ^{2}}{(n-\frac{1}{2})\gamma 
s\ln(\lambda/r_{w})},\; E_{\times} =-\frac{A\Phi _{0}^{2}}{4\pi 
^{2}\gamma ^{2}s\ln(A_{1}\gamma s/\lambda)}.
\]
Numerical calculations give $A=\sum_{n=1}^{\infty 
}\left(\frac{C_{n}}{n-1/2}\right)^{2}\approx 2.1$ and $A_1\approx 3.5$. 
At finite $B_{x}$ each PV stack intersects $1/c_{z}$  JVs per unit 
length and the interaction energy of the PV stack with the JV stack is 
given by 
\begin{equation} 
\varepsilon_{Jp}=\frac{E_{\times}}{c_{z}}=\sqrt{\frac{\sqrt{3}\gamma 
B_{x}}{2 \Phi _{0}}}\frac{A\Phi _{0}^{2}}{4\pi ^{2}\gamma 
^{2}s\ln(A_1\gamma s/\lambda)}
\label{Eint} 
\end{equation} 
The transition to the ``chain state'' is expected when the energy gain 
(\ref{Eint}) to put an extra PV stack on the JV stack exceeds the 
energy loss due to formation of the edge interstitial in the Abrikosov 
lattice $\varepsilon _{EI}$ at $a\gg \lambda$
%recently calculated by Olive and Brandt 
\cite{Olive}, $\varepsilon _{EI}\approx 1.14\varepsilon 
_{0}\sqrt{a/\lambda }\exp ( -a/ \lambda )$. 
%\sqrt{\frac{a}{\lambda }}\exp \left( -\frac{a}{ \lambda }\right)$.  
Comparison of the two energies gives the following equation for the PV 
lattice spacing $a_{t}$ at which the transition takes place
\begin{equation} 
\sqrt{\frac{a_{t}}{\lambda }}\exp \left( -\frac{a_{t}}{\lambda 
}\right) \approx \sqrt{\frac{\gamma s^{2}B_{x}}{\Phi _{0}}}\frac{ 6.86 
\lambda ^{2}}{\gamma ^{2}s^{2}\ln(3.5\gamma s/\lambda)}
\label{at}
\end{equation} 
 The transition field $B_{t}=2\Phi_{0}/\sqrt{3}a_{t}^{2}$ grows slowly with 
 $B_{x}$.  For $\gamma=200$, $\lambda =220$nm, and $B_x=30$G we obtain 
 $B_{t}\approx 36$G. The transition can be driven by the field or by the 
 temperature (due to the T-dependence of $\lambda$).

In conclusion, we have shown that in very anisotropic layered 
superconductors the vortex ground state in tilted fields is given by the 
crossing lattices of the JVs and PV stacks.  This ground state explains 
(i)the linear dependence of the c-component of the melting field on the 
in-plane field and (ii)the mixed chains-lattice state observed by Bitter 
decorations at small fields.

I am very grateful to M.~Konczykowski for providing the data of 
Ref.~\onlinecite{Schmidt}.  I would like also to thank L.~N.~Bulaevskii and 
W.~K.~Kwok for useful comments.  This work was supported by the NSF Office 
of the Science and Technology Center under contract No.~DMR-91-20000 and by 
the U.~S.\ DOE, BES-Materials Sciences, under contract 
No.~W-31-109-ENG-38.
%\vspace*{-10pt}
%\begin{references}

%\end{references}
\vspace*{-10pt}
\vspace*{-10pt}
\end{document}